\documentclass[twocolumn]{revtex4}

\input{epsf.tex}

\newcommand{\s}{\sigma}

\renewcommand{\a}{\alpha}

\newcommand{\be}{\begin{equation}}
\newcommand{\ee}{\end{equation}}
\newcommand{\bea}{\begin{eqnarray}}
\newcommand{\eea}{\end{eqnarray}}
\newcommand{\ba}{\begin{array}}
\newcommand{\ea}{\end{array}}
\def\J#1#2#3#4{{#1} {#2} (#3) #4}
\def\PRD{Phys. Rev. D}
\def\PR{Phys. Rev.}
\def\PRL{Phys. Rev. Lett.}

\def\PTP{Prog. Theor. Phys.}

\def\JMP{J. Math. Phys.}

\def\CQG{Class. Quantum Grav.}

\def\PLB{Phys. Lett. B}

\def\GC{{Gravit. Cosmology}}

\begin{document}
\draft
\title{Comments on two papers of Cl\'ement and Gal'tsov}

\author{H. Garc\'ia-Compe\'an$^a$, V.S. Manko$^a$, E. Ruiz$^b$}
\address{$^a$Departamento de F\'\i sica, Centro de Investigaci\'on y de
Estudios Avanzados del IPN, A.P. 14-740, 07000 Ciudad de M\'exico,
Mexico\\$^b$Instituto Universitario de F\'{i}sica Fundamental y
Matem\'aticas, Universidad de Salamanca, 37008 Salamanca, Spain}

\begin{abstract}
We comment on physical inconsistences of the Cl\'ement-Gal'tsov
approach to Smarr's mass formula in the presence of magnetic
charge. We also point out that the results of Cl\'ement and
Gal'tsov involving the NUT parameter are essentially based on the
known study (dating back to 2006) of the Demia\'nski-Newman
solutions which was not cited by them.
\end{abstract}


\maketitle


In the paper \cite{CGa}, Cl\'ement and Gal'tsov considered the
mass and angular momentum distributions in the dyonic Kerr-Newman
(KN) black-hole spacetime \cite{KNe,Car} to get the results
different from those earlier obtained for this spacetime in
\cite{MGa}. The preprint \cite{MGa} was later published under a
slightly different title \cite{MGa2} better reflecting the topic
of the special issue of Classical and Quantum Gravity on black
holes and electromagnetic fields, and the paper \cite{CGa} was not
mentioned there because the physical inconsistences in the
formulas (4.8) and (4.14) of \cite{CGa} were so glaring, that we
hoped Cl\'ement and Gal'tsov would be able to detect these
themselves. However, it appears that the aforementioned authors
were pretty sure about the correctness of their results because in
the recent paper \cite{CGa2} they have extended their approach
further to the solutions with the NUT parameter \cite{NTU},
hinting in passing that the title change of the preprint
\cite{MGa} might have had something to do with the critical tone
of their previous work \cite{CGa}. Therefore, we now feel
ourselves obliged to respond the Cl\'ement and Gal'tsov's
critique, and in what follows we will comment on the physical
inconsistences of the papers \cite{CGa,CGa2}; moreover, we will
also point out a research article whose results have been
appreciably used (but not cited) in the paper \cite{CGa2}.

We start by noting that in \cite{MGa} it was shown how the
magnetic charge can be elegantly introduced into the well-known
Smarr mass formula \cite{Sma} for black holes, and the extended
formula was applied to several dyonic black-hole systems. The
paper \cite{CGa} of Cl\'ement and Gal'tsov addresses a technical
issue of the evaluation of mass by arguing that Tomimatsu's mass
integral \cite{Tom} that was employed in \cite{MGa} must have an
additional term affecting the distribution of mass along the
symmetry axis. In the case of the dyonic KN solution, for example,
the appearance of the new term leads, according to \cite{CGa}, to
an exotic model with three massive regions on the symmetry axis --
the central one with mass $M_H$ and two semi-infinite strings of
masses $M_{S_\pm}$ attached to the central region (see Fig.~1) --
and for $M_H$ and $M_{S_\pm}$ Cl\'ement and Gal'tsov obtained the
following analytical expressions \cite{CGa}:
\bea &&M_H=M-\frac{P^2(M+\s)}{(M+\s)^2+a^2}, \nonumber\\
&&M_{S_\pm}=\frac{P[P(M+\s)\mp a Q]}{2[(M+\s)^2+a^2]}, \nonumber\\
&&\s=\sqrt{M^2-a^2-Q^2-P^2}, \label{Mpm} \eea
which satisfy the relation $M_H+M_{S_+}+M_{S_-}=M$, where $M$
stands for the total mass (the remaining parameters $a$, $Q$ and
$P$ are, respectively, the ratio of the total angular momentum $J$
and total mass $M$, the electric charge and the magnetic charge).

\begin{figure}[htb]
\centerline{\epsfysize=70mm\epsffile{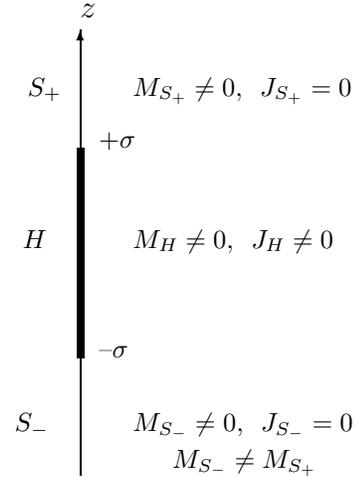}}
\caption{Distribution of mass and angular momentum along the
symmetry axis in the Cl\'ement-Gal'tsov model of the dyonic KN
solution. The three different parts of the symmetry $z$-axis are:
$z>\s$ (part $S_+$, the upper part of the axis), $-\s<z<\s$ (part
$H$, the horizon) and $z<-\s$ (part $S_-$, the lower part of the
axis).}
\end{figure}

A simple inspection of formulas (\ref{Mpm}) in the subextreme case
($M^2>a^2+Q^2+P^2$), however, reveals that the model proposed and
advocated by Cl\'ement and Gal'tsov as alternative to the usual
interpretation of $M$ (the mass fully confined inside the central
body) has several frankly {\it unphysical features}. First, the
semi-infinite strings introduced in \cite{CGa} have different
masses $M_{S_\pm}$, which apparently contradicts the {\it
equatorial symmetry} of the dyonic KN solution (see \cite{EMR,PSa}
for the definition of the equatorially symmetric electrovac
spacetimes) requiring $M_{S_+}=M_{S_-}$. Moreover, it is easy to
see that for small values of the magnetic charge $P$ the masses
$M_{S_+}$ and $M_{S_-}$ of the two strings can even take opposite
signs, which introduces undesirable {\it negative} masses into a
well-behaved solution. Mention also that the parameter $a$ in the
Cl\'ement-Gal'tsov treatment does not represent the total angular
momentum per unit mass calculated on the horizon because the parts
$S_\pm$ of the symmetry axis have zero angular momenta and nonzero
masses, thus contradicting Carter's interpretation \cite{Car} of
the dyonic KN solution.

There are several possible explanations for the origin of the
physically unrealistic formulas (1). At the first try, the
appearance of the additional term in the mass integral (3.11) of
\cite{CGa} leading to the above (1) could be attributed to the
clearly erroneous equations (3.2) of \cite{CGa} defining the
magnetic scalar potential $u$ ($A^{'}_\varphi$  in the notation of
\cite{MGa2}). At the same time, even if the calculations of
Cl\'ement and Gal'tsov are somehow correct, the presence of the
term involving the product $A_\varphi u$, $A_\varphi$ being the
magnetic component of the electromagnetic 4-potential, must not
really produce any effect on the usual physical interpretation of
the dyonic KN solution because there are arguments in favor of
{\it vanishing} of such a term. Indeed, taking into account that
the potential $A_\varphi$ of a magnetic dipole vanishes on the
$S_\pm$ parts of the symmetry axis, one naturally comes to the
idea that in the case of a magnetic monopole the respective
$A_\varphi$ can also be made equal to zero on $S_\pm$ if one
treats the Dirac string as a ``gauge artifact'' \cite{CLi}, which
allows for choosing an appropriate value of the integration
constant $b_0$ in the expression of $A_\varphi$ on each part of
the symmetry axis. Then the potential $A_\varphi$ of the dyonic KN
solution, namely,
\be A_\varphi=b_0-Py-a(1-y^2)A_t, \label{Aphi} \ee
where $A_t$ is the electric potential and $y$ the ellipsoidal
coordinate, will take zero value on $S_+$ ($y=+1$) after choosing
$b_0=P$, while on the lower part of the symmetry axis $S_-$
($y=-1$) the potential $A_\varphi$ vanishes at $b_0=-P$.
Consequently, in this case both $M_{S_+}$ and $M_{S_-}$ also
become zeros, which is consistent with the regularity of the
metric on $S_\pm$. Obviously, this approach is equivalent to
calculating $M_{S_\pm}$ (and $M_H$ too) by means of the usual
Tomimatsu's mass integral.

Furthermore, it is worth noting (setting aside the quantum aspects
of magnetic monopoles) that classically the electric and magnetic
charges are expected to exhibit similar properties \cite{Jac}, in
particular with respect to the singularity structure of their
physical fields. In general relativity, within the framework of
Ernst's formalism of complex potentials \cite{Ern}, this
similarity manifests itself through the invariance of the Ernst
equations under the duality rotation of the electromagnetic
potential $\Phi\to e^{i\a}\Phi$, $\a={\rm const}$, so that the
same metrics can describe geometries induced either by an electric
or magnetic charge, or by both. A good evidence of similarity
among the two charges is provided by the dyonic
Reissner-Nordstr\"om solution \cite{Car}, for which the energy
density of the electromagnetic field $-T_t^t$ can be shown to have
the form
\be \frac{Q^2+P^2}{8\pi r^4}, \label{den} \ee
$r$ being the radial coordinate, and one can see that the magnetic
charge contributes into the electromagnetic energy on an equal
footing with the electric charge. Moreover, the electric and
magnetic charges of the dyonic KN solution are both located inside
the horizon, so we see no plausible physical reasons to consider
that they must affect differently the distribution of mass in the
solution.

It should be also pointed out that, while constructing exact
solutions, a proper choice of the integration constants is of
paramount importance for the correct physical interpretation of
the solutions. In the stationary vacuum case, at least two metric
functions are defined up to additive constants, the choice of
which is determined by the boundary conditions, and it is
precisely for the physical reasons, say, the Kerr metric
\cite{Ker} has only two arbitrary real parameters instead of four.
In the case of stationary electrovac spacetimes, an additional
integration constant may arise in the expression of the
electromagnetic potential, and it is clear that its choice must be
congruent with the geometrical and physical properties of the
metric. Apparently, in the paper \cite{CGa}, Cl\'ement and
Gal'tsov were unable to resolve a rather nontrivial and subtle
problem of the parameter choice in the potential $A_\varphi$ in
the presence of spurious singularities, and they elaborated and
presented an absolutely weird interpretation of the dyonic KN
black-hole solution, which can hardly be justified even by the yet
hypothetical status of magnetic charges.

In the subsequent paper \cite{CGa2}, Cl\'ement and Gal'tsov
extended their specific ideas about the magnetic charge to a NUT
generalization of the dyonic KN solution. The nonzero NUT
parameter endows the metric with a pair of semi-infinite
singularities located on the symmetry axis which, in
contradistinction from the fictitious singularities of the
potential $A_\varphi$ describing the magnetic monopole, do affect
the mass distribution in the solution. Here, it would be
worthwhile noting that the first study of the physical properties
of the NUT singularity was undertaken by Bonnor \cite{Bon} with
the aid of an approximation method, and he interpreted it as a
{\it massless} source of {\it finite} angular momentum. This
actually erroneous interpretation was rectified only 36 years
later in the paper \cite{MRu}, where it was rigorously proven that
the NUT singularity is {\it massive} and carries {\it infinite}
angular momentum; besides, there exists a unique choice of the
integration constant at which the total angular momentum of the
NUT solution takes finite (zero) value, and it corresponds to the
case of two counter-rotating semi-infinite singularities attached
to the nonrotating central body. Later, the properties of the NUT
singularities in the more general metrics were also analyzed
\cite{MMR}, and in this respect we would like to mention that the
so-called Kerr-NUT and dyonic Kerr-Newman-NUT solutions were both
obtained for the first time by Demia\'nski and Newman \cite{DNe},
who also gave the name `Kerr-NUT' to their vacuum spacetime.

Since the original form of the Demia\'nski-Newman (DN) 5-parameter
electrovac solution is not quite suitable for applications, in the
papers \cite{MMR,AGM} another representation of the DN metric was
worked out within the framework of the extended $N$-soliton
electrovac spacetime \cite{RMM}. In \cite{MMR} the choice of the
integration constant at which the two DN solutions have finite
angular momentum was established, and the distributions of mass
and angular momentum along the symmetry axis were studied
separately in the vacuum and electrovac cases. Surprisingly, in
the recent paper \cite{CGa2}, Cl\'ement and Gal'tsov have
presented a fairly similar analysis of the mass and angular
momentum distributions in the Kerr-NUT (vacuum DN) and dyonic
KN-NUT (electrovac DN) solutions, and for their purpose they made
use of the representations obtained in the paper \cite{MMR} for
the DN spacetimes. However, in their paper they do not give any
reference on the work \cite{MMR}, neither they cite the original
paper \cite{DNe} of Demia\'nski and Newman where the solutions
were first constructed. With regard to the electrovac DN solution
we would only like to point out that the basic formulas
(3.52)-(3.55) of \cite{CGa2} are precisely formulas (3), (10) of
\cite{MMR} in which Cl\'ement and Gal'tsov performed the following
formal redefinitions of the parameters:
\bea &&a\to-a, \quad \nu\to n, \quad q\to-q, \quad b\to p,
\nonumber\\ &&C_1\to 0, \quad C_2\to 0. \label{red} \eea
Apparently, the results of Cl\'ement and Gal'tsov involving the
magnetic charge and NUT parameter are plagued with the same
problems as already discussed earlier in the case of the dyonic KN
solution, so no further comments on that are really needed.

As far as the Kerr-NUT solution is concerned, the paper
\cite{CGa2} deserves special remarks to be made. First, it is very
clear that the form (3.40) of the Kerr-NUT metric given in
\cite{CGa2} is {\it identical} with the form (10) of \cite{MMR};
of course, (3.40) was not obtained from (2.9)-(2.11) of
\cite{CGa2} by two successive coordinate transformations, as
affirmed by Cl\'ement and Gal'tsov, but rather by just setting to
zero the charge parameters $q$ and $p$ in the electrovac DN
(dyonic KN-NUT) solution, like this was also done in \cite{MMR}.
Moreover, the fact that the coordinates $x$ and $y$ are
erroneously called in \cite{CGa2} the {\it prolate} spheroidal
coordinates is a clear indication that Cl\'ement and Gal'tsov do
not really understand the notion of the {\it generalized}
spheroidal coordinates $x$ and $y$ introduced for the DN solutions
in \cite{AGM} to cover both the {\it real} and {\it pure
imaginary} sectors of the quantity
$\s=\sqrt{m^2+n^2-a^2-q^2-p^2}$, where all five parameters can
take arbitrary real values. In the usual prolate spheroidal
coordinates, $\s$ appears as an arbitrary {\it real} parameter.

Second, all the formulas given in subsection 3.3 of \cite{CGa2}
for the distributions of mass and angular momentum in the Kerr-NUT
spacetime had already been obtained in section 3 of \cite{MMR}
even for arbitrary values of the integration constant $C$ ($s$ in
\cite{CGa2}) entering the expression of the metric function
$\omega$. An important physical message of the paper \cite{MMR}
was that the NUT parameter always introduces {\it negative} mass
via the semi-infinite singularities.

Third, Cl\'ement and Gal'tsov consider in \cite{CGa2} what they
call ``a {\it symmetric} Misner string configuration'',
corresponding to $s=0$ by analogy with the pure NUT case
\cite{MRu}. However, unlike in the latter case, in the generic
Kerr-NUT solution there is no any ``symmetric'' configuration of
two semi-infinite singularities at any $s$ because the Kerr
rotational parameter $a$ introduces the asymmetry into the
combined Kerr-NUT metric (due to the counter-rotation of strings).
Therefore, in principle the choice $s=0$ leading to the finite
angular momentum in the Kerr-NUT solution needs a rigorous
justification, which was actually done in \cite{MMR}, and the
unequal masses of the two semi-infinite singularities in this case
clearly show that the singularity structure defined by $s=0$ is
{\it asymmetric} indeed. Moreover, as was shown in \cite{MMR}, the
aggregate mass of the two singularities in the $s=0$ configuration
is a negative quantity, which invalidates, in our opinion, the
importance of the Kerr-NUT solution for thermodynamics.

Lastly, the results on the physical properties of the DN vacuum
and electrovac solutions obtained in \cite{MMR} significantly
improve the understanding of the NUTty spacetimes, and we find it
quite regretful and unfair that Cl\'ement and Gal'tsov not only
attempted to ascribe to themselves the most important findings of
the paper \cite{MMR}, but also gave erroneous statements about
some questions well clarified nearly fifteen years ago.

\section*{Acknowledgments}

One of the authors (VSM) is indebted to Brandon Carter for
interesting correspondence on magnetic charges in 2005. This work
was partially supported by Project~128761 from CONACyT of Mexico,
by Project PGC2018-096038-B-100 from Ministerio de Ciencia,
Innovaci\'on y Universidades of Spain, and by Project SA083P17
from Junta de Castilla y Le\'on of Spain.

\end{document}